\documentclass[aps,prl,amsmath,amssymb,onecolumn,nofootinbib,showpacs,tightenlines,12pt]{revtex4}

\usepackage{epsfig,graphicx}
\usepackage{bm}
\newcommand{\be}{\begin{equation}}
\newcommand{\ee}{\end{equation}}
\newcommand{\bea}{\begin{eqnarray}}
\newcommand{\eea}{\end{eqnarray}}

\begin{document}
\title{Phase transition in Schwarzschild-de Sitter  spacetime }
\author{D.Momeni}
\email{davood.momeni@kiau.ac.ir} \affiliation{
 Department of Physics ,Faculty of science,Islamic Azad University, Karaj Branch. .Iran, Karaj,
Rajaei shahr, P.O.Box: 31485-313 }
\author{A.Azadi}
\email{azadi@phymail.ut.ac.ir} \affiliation{Department of Physics,
University of Tehran, North Karegar Ave., 14395-547, Teheran, Iran }

\begin{abstract}
  Using a static massive spherically symmetric scalar field coupled to gravity in the
Schwarzschild-de Sitter (SdS) background, first we consider some
asymptotic solutions near horizon and their local equations of
state(E.O.S) on them. We show that near cosmological and event
horizons our scalar field behaves as a dust. At the next step near
two pure de-Sitter or Schwarzschild horizons we obtain a coupling
dependent pressure to energy density ratio. In the case of   a
minimally couplling this ratio is $-1$ which springs to the mind
thermodynamical behavior of dark energy. If having a negative
pressure behavior near these horizons we concluded that the coupling
constant must be $\xi<\frac{1}{4}$ . Therefore we derive a new
constraint on the value of our coupling $\xi$ . These two different
behaviors of unique matter in the distinct regions of spacetime at
present era can be interpreted as a phase transition from dark
matter to dark energy in the cosmic scales and construct a unified
scenario.
\end{abstract} \maketitle
\section{\label{sec:level1}Introduction}
After many years of considering that the Universe is filled with a
dark matter and with a dark energy causing an acceleration of the
expansion of the Universe, we have still no direct evidence of their
existences. In addition , observations of supernova of type Ia tend
to cause trouble to common dark energy models \cite{1}, and open the
way to new kind of models to explain the observed repulsing effects
\cite{2}. Alternative models has created to explain the cosmological
observations, and in particular some of them try to solve the dark
energy and dark matter problems by unified object in which the dark
matter and the dark energy are in fact different aspects of a same
fluid, the "dark fluid". For instance mentioned that the Generalized
Chaplygin Gas model \cite{3} follows this idea and is presently
under scrutiny. .As scalar field-based models for dark energy and
dark matter exist in the literature \cite{4}, it seems interesting
to investigate unifying dark fluid models based on scalar fields.
This idea has been proposed in \cite{5}.

On the other hand spherical symmetric solutions of a  scalar field
in static spacetimes is the most simple case to study \cite{6}. We
know that there is no simple exact solution for a static massive
scalar field as a source of gravity.  Some attempts to construct
such model were failed because of mathematically complications.
  There is only a massless minimally coupled scalar field version of
gravity  \cite{7} . If we insert mass or coupling constant, equation
can not be integrated in closed forms. Considering unverse
acceleration, if we assume that our universe be asymptotically de
Sitter and there exists a massive black hole in the center of
galaxy,for the metric we can write a simple schwarzschild-de Sitter
line element which in weak field approximation asymptotically leads
to de Sitter.There is some  significant works on the SdS black hole
from stability viewpoint of higher dimensional spacetimes \cite{9}
and also on calculation of quasinormal modes of four dimensional
SdS black holes  \cite{10}. In this work we
 present a massive-coupled-scalar-field in the SdS background
and show that in the scale of our cosmos,  dust and negative
pressure matter are the same and essentially can be unified in
present era as a negative pressure matter that evolves like a dust
near the horizons.
\section{Sds geometry}
  SdS space time is a well known vacuum solution of   Einstein field
equations with a constant curvature   $R=+\frac{12}{a^{2}}$  where
$a$ is related to Cosmological-Constant through
$\Lambda=\frac{3}{a^{2}}$  . It's geometry as a  4-dim Riemmanian
manifold can be represented as:
\begin{eqnarray}
ds^{2}=-f(r)dt^{2}+1/f(r)dr^{2}+r^{2}(d\theta^{2}+sin^{2}\theta
d\varphi^{2})
\end{eqnarray}
 Where
 $f(r)=1-\frac{2M}{r}-\frac{r^2}{a^2}$ . Depending on the values
 of  $9M^2\Lambda$  where $M$ is black-hole (B.H.) mass we have three
 distinct cases:\\
 1. If   $9M^2\Lambda=1$    we have an extremal B.H. In this
case both cosmological and  event horizons coincide. That
is $r_{+}=r_{c}$. \\
2. If $9M^2\Lambda>1$
 We've no B.H.  solution .\\
3. If   $0<9M^2\Lambda<1$   we have  Black hole  solution with two
positive horizons as cosmological(which is denoted by $r_{c}$) and
event horizons $r_{+}$ and an un-physical negative horizon as:
 $r_{n}=-(r_{c}+r_{+})$ . In this work we only restricted ourselves to
Black hole  case.
\section{Field Eqs}
 A simple coupled massive scalar field in a curved
 space-time  is represented by a Lagrangian \cite{7}:
\begin{eqnarray}
\label{aa}
 L=\frac{1}{2}(\phi_{;\mu}\phi^{;\mu}-\xi R
\phi^{2}-m^{2}\phi^{2})
\end{eqnarray}
 Matter action is:
\begin{eqnarray}
\label{aa}
 S_{matter}=\int \sqrt{-g}\frac{1}{2}(\phi_{;\mu}\phi^{;\mu}-\xi R
\phi^{2}-m^{2}\phi^{2})d^{4}x
\end{eqnarray}
Related Energy-Momentum
  tensor and  equation of motion(E.O.M) for field are \cite{8} :
\begin{eqnarray}
T_{\mu\nu}=(1-2\xi)\phi_{;\mu}\phi_{;\nu}+(2\xi-\frac{1}{2})g_{\mu\nu}
(g^{\rho\sigma}\phi_{;\rho}\phi_{;\sigma})&&\\\nonumber-2\xi\phi_{;\mu\nu}\phi
+\gamma g_{\mu\nu}\phi^{2}\\\nabla_{\mu}\nabla^{\mu}\phi+(m^{2}+\xi
R)\phi=0
\end{eqnarray}
  In a SdS  space time  we denote $\gamma$ :
\begin{eqnarray}
\label{aa}
 \gamma=m^{2}\frac{1-4\xi}{2}-\frac{3\xi(8\xi+1)}{a^{2}}
\end{eqnarray}
  Where m is our scalar field mass. For a SdS B.H  we can take scalar
field as a function of radial coordinate $r$ to preserve isotropy
 of   spacetime . Also we can ignore  from all
back reaction effects  in this model. The energy density and
pressure are:
\begin{eqnarray}
\rho=g^{00}T_{00}=f(r)(2\xi-\frac{1}{2})(\phi^{'})^{2}-\xi
f(r)^{'}\phi\phi^{'}+\gamma\phi^{2} \\
P_{r}=g^{rr}T_{rr}
=\frac{1}{2}f(r)(\phi^{'})^{2}-\nonumber2\xi\phi(f(r)\phi^{''}\\&&\hspace{-50mm}+\frac{1}{2}(f(r))^{'}\phi^{'})+\gamma\phi^{2}
\end{eqnarray}
 For a massless-minimally coupled scalar field our expressions for
radial pressure and energy density become very simple :
\begin{eqnarray}
\rho=-\frac{1}{2}f(r)(\phi^{'})^{2}\\
P_{r}=\frac{1}{2}f(r)(\phi^{'})^{2}
\end{eqnarray}
 Also for a massive minimally coupled model we have :
\begin{eqnarray}
\rho=-\frac{1}{2}f(r)(\phi^{'})^{2}+\frac{1}{2}m^{2}\phi^{2}
\nonumber\\&&\hspace{-50mm}
 P_{r}=\frac{1}{2}f(r)(\phi^{'})^{2}+\frac{1}{2}m^{2}\phi^{2}
\nonumber\\&&\hspace{-50mm}
\end{eqnarray}
  Field equation of motion is :
\begin{eqnarray}
\frac{1}{r^{2}}\frac{d( r^{2} f(r)\frac{d \phi}{ d r})}{d
r}+\mu^{2}\phi=0
\end{eqnarray}
  Where :
\begin{eqnarray}
  \mu^{2}=\xi R+m^{2}
\end{eqnarray}
  Also we define index $w$ as :
\begin{eqnarray}
w=\frac{P}{\rho}
\end{eqnarray}
 In this equation pressure is radial  component of
energy-momentum tensor. We choose radial pressure to preserve
isotropy of our spacetime and not to disturb the matter distribution
enclosed in our model.
\section{Asymptotic solutions}
 In  equation   $(12)$  we
 substitute $\phi(r)$ by $\psi(r)$ :
\begin{eqnarray}
\phi(r)=\frac{\psi(r)}{r\sqrt{f(r)}},
\end{eqnarray}
 Which gives :
 \begin{eqnarray}
 \frac{d^{2}\psi(r)}{d
r^{2}}+v(r)\psi(r)=0
\end{eqnarray}
 Where :
\begin{eqnarray}
v(r)=\frac{1}{4}(\frac{ f(r)'}{f(r)})^{2}-\frac{f(r)'}{r
f(r)}-\frac{1}{2}\frac{f(r)''}{f(r)}+\frac{\mu^{2}}{f(r)}
\end{eqnarray}
  We mention here that there  is no orthogonal basis functions
for this equation and we can not  construct an  exact solution for
this equation  and one can show that only there is an asymptotic
semi-analytic solution for it. In fact if we use series method we
take :
\begin{eqnarray}
\psi(r)=\sum _{n=6}^{\infty }b _{ n } {r}^{n}
\end{eqnarray}
 Where recursion relation for series coefficients is :
\begin{eqnarray}
&           &4\,b_{{n}} \left( x+y \right) ^{2} \left( n-1/2 \right)
^{2}{y}^{2}{x} ^{2}-\\\nonumber&&8\,b_{{n-1}}xy \left( n-1 \right)
\left( n-2 \right)  \left( x+y \right)  \left( {y}^{2}+xy+{x}^{2}
\right)
\\\nonumber&&+4\,b_{{n-2}} \left( n-2 \right)  \left( n-3 \right)  \left(
{y}^{2}+xy+{x}^{2} \right) ^{2}\\\nonumber &&-4 \,b_{{n-3}}xy \left(
{\mu}^{2}{a}^{2}-2\,{n}^{2}+14\,n-20 \right) \left( x+y
\right)\\\nonumber &&+4\,b_{{n-4}} \left( {y}^{2}+xy+{x}^{2} \right)
\left( -2\,{n}^{2}+18\,n-37+{\mu}^{2}{a}^{2} \right)
\\\nonumber&&+b_{{n-6}}
 \left( 160+4\,{n}^{2}-52\,n-4\,{\mu}^{2}{a}^{2}
 \right)=0,n\geqslant 6
\end{eqnarray}
 There is no general method to solving this equation and  we
 can use from an iteration method based on this equation which we
 do not present it here because it is difficult to represent  coefficients in terms of polynomials.
  But we can prove that since the usual convergence tests for infinite series is valid in the  interval of  $x<r<y$  the series solution (18)  is
 converged.\\
 In fact if we denote coefficients of
  \begin{eqnarray}
\nonumber&&b_{n-k},k=0,...6
 \end{eqnarray}
  by $c_{k}$ then we must show that
   \begin{eqnarray}
   \nonumber&&
\lim_{n\rightarrow \infty}\frac{b_{n+1}}{b_{n}}=\mathfrak{L}<1
\end{eqnarray}
This inequality must be valid for all terms of
$b_{n}$   thus we must have
\begin{eqnarray}
\nonumber&& \lim_{n\rightarrow
\infty}\frac{b_{n-k}}{b_{n-k+1}}=\mathfrak{L}<1,\forall
k=2,3,4,6
\end{eqnarray}
 But it is obvious that:
 \begin{eqnarray}
\nonumber&&  \lim_{n\rightarrow
\infty}\frac{b_{n+1}}{b_{n}}<-\frac{c_{1}+c_{2}+c_{3}+c_{4}+c_{6}}{c_{0}}
\end{eqnarray}\\
 We define a new function
\begin{eqnarray}
\nonumber&&
 F=-\frac{c_{1}+c_{2}+c_{3}+c_{4}+c_{6}}{c_{0}}
 \end{eqnarray}.
  In terms of  x,y
 explicitly we  can write that :\\
\begin{eqnarray}
\nonumber&&
F-1=-\frac{(y-1)^{2}(x-1)^{2}(y+x+1)^{2}}{(y+x)^{2}(y)^{2}(x)^{2}}<0
\end{eqnarray}
 Which is hold for all values of   x,y  also  if  $y>x$. Then series
 (18) is converged. Now
  we take one simple  approach to the problem. The best approach
 is using asymptotic expansion of solutions near  physical horizons
  $r_{c},r_{+}$   and considering in details the phase transition .
  For simplicity we denote horizons   $r_{c}$   by   $x$
and  $r_{+}$  by $y$ . Then by simple algebra we have  :
\begin{eqnarray}
f(r)=\frac{-(r-x)(r-y)(r+x+y)}{a^{2} r}
\end{eqnarray}
 By expanding $v(r)$ near $r=x$ we obtain :
\begin{eqnarray}
v(r)\simeq \frac{1}{4}(r-x)^{-2}+O[(r-x)^{-1}]
\end{eqnarray}
Then near $r=x$  asymptotic solution  for (16) is :
\begin{eqnarray}
\psi(r)=\sqrt{r-x}(c_{1}+c_{2}{\rm ln}{(r-x)})
\end{eqnarray}
  This solution vanishes identically near $r=x$ and satisfy
Dirichlet-Boundary-Condition(D.B.C) on the surface  $r=x$ . By
rewriting this solution in the form of $\phi(r)$ we obtain :
\begin{eqnarray}
\phi(r)=\frac{\sqrt{r-x}(c_{1}+c_{2}{\rm ln}{(r-x)})}{r\sqrt{f(r)}}
\end{eqnarray}
 Thus near $r=x$, the expansion  of E.O.S (14) up to 2'nd order is :
\begin{eqnarray}
w\simeq 0
\end{eqnarray}
 Similarly  near $r=y$ :
\begin{eqnarray}
v(r)\simeq \frac{1}{4}(r-y)^{-2}+O[(r-y)^{-1}]
\end{eqnarray}
 With the same  D.B.C we've :
\begin{eqnarray}
\psi(r)=\sqrt{r-y}(c_{1}+c_{2}{\rm ln}{(r-y)})
\end{eqnarray}
 Thus near $r=y$, E.O.S up to 2'nd order is :
\begin{eqnarray}
w\simeq 0
\end{eqnarray}
 Thus our scalar field near both horizons behaves like a dust
with    $w=0$.\\
On the other hand spacetime which close to the B.H. is
Schwarzschild-like and far away from $y$ we have de Sitter spacetime
with no singularity therefore we could preserve the form (1) .Near
both Schwarzschild or de Sitter horizons(i.e; $r=2M$ $\&$$r=a$ )
solutions for (16)  are derived from a simple Airy function with a
simple D.B.C as mentioned before that is :
\begin{eqnarray}
\psi(r)=-AiryA(\frac{v_{0}+v_{1}(r-h)}{-v_{1}^{2/3}})AiryB
(\frac{-v_{0}}{v_{1}^{2/3}})\nonumber\\&&\hspace{-77mm}+AiryB(\frac{v_{0}+v_{1}(r-h)}{-v_{1}^{2/3}})AiryA
(\frac{-v_{0}}{v_{1}^{2/3}})
\end{eqnarray}
   Where $h$ can be $2M$ or $a$, $v_{0}=v(r=h)$
and $v_{1}=v'(r=h)$. Finally near this region we've :
\begin{eqnarray}
w\simeq(\frac{1}{4\xi-1})+O[(r-h)^{2}]
\end{eqnarray}
 Which for a minimally coupled scalar field our matter behaves like
 an exotic matter with a similar equation of state as Dark energy in
 the cosmos does. If we  focused only on the negative pressure
 matters we obtain a new fascinated constraint on the value of
 coupling as :
\begin{eqnarray}
\xi<\frac{1}{4}
\end{eqnarray}
 Which   both interesting cases (minimally and conformally
 coupled)belongs to this interval. We mention here that this idea
, scalar fields can  change a phase  from dust to negative
 pressure matter is an anstaz and we can not deduce a full complete
description of the mechanism which is hidden behind this event.
\section{some further physical notes}
 Why we take scalar field responsible for phase transition? This is an
 anstaz for simplicity. We should mention here that there is no exact
 solution for a spherically symmetric solution for field equations
 with a massive non minimally coupled source . Thus
 from a mathematically point of view, one can take scalar field  as a
 perturbation to our spherically symmetric metric in four dimensions. Also it can be
 shown that a minimally coupled -massless -scalar field in any
 spherically symmetric spacetime behaves like a fluid with equation
 of state $P=-\rho$. That is dark energy!. Thus any
 minimally coupled scalar field in a spherically symmetric background
 behaves like a pure dark energy. One can show that there is no
 phase transition from dark energy to dark matter in this case.
 Therefore phase transition is caused just for none minimally coupled
 massive scalar field. Therefor the local E.O.S. doesn't include
  mass ($m^2$) to first order so even having a coupled
  massless scalar field could cause this transition. \\

\section{summary}
 In this work first we derived some asymptotic solutions near both
 cosmological and event horizons of SdS and showed that our field behaves as
 a dust (dark matter) near these
 horizons. We showed that far a way from them our field behaves like a
 matter with negative pressure and this limited ourselves to
 some constraints about the coupling constant $\xi$.
\section{Acknowledgement}
 This work was supported by University of Tehran (Iran)

\end{document}